\documentclass[aps,pra,preprintnumbers,showpacs,tightenlines]{revtex4}
\usepackage{amssymb}
\usepackage{amsmath}
\usepackage{graphicx}
\usepackage{epsfig}
\usepackage{subfigure}
\usepackage{amsfonts}
\usepackage{CJK}

\begin{document}

\title{Single-step implementation of a hybrid controlled-NOT gate with one superconducting qubit simultaneously controlling multiple
target cat-state qubits}

\author{Qi-Ping Su$^{1}$}
\author{Yu Zhang$^{2}$}
\author{Chui-Ping Yang$^{1,3}$}
\email{yangcp@hznu.edu.cn}

\address{$^1$Department of Physics, Hangzhou Normal University, Hangzhou 311121, China }
\address{$^2$School of Physics, Nanjing University, Nanjing 210093, China}
\address{$^3$Quantum Information Research Center, Shangrao Normal University, Shangrao 334001, China}

\date{\today}

\begin{abstract}
Hybrid quantum gates have recently drawn considerable attention. They play significant roles
in connecting quantum information processors with qubits of different encoding and have
important applications in the transmission of quantum states between a quantum processor
and a quantum memory. In this work, we propose a single-step implementation of a multi-target-qubit
controlled-NOT gate with one superconducting (SC) qubit simultaneously controlling $n$
target cat-state qubits. In this proposal, the gate is implemented with $n$
microwave cavities coupled to a three-level SC qutrit. The two logic states
of the control SC qubit are represented by the two lowest levels of the
qutrit, while the two logic states of each target cat-state qubit are
represented by two quasi-orthogonal cat states of a microwave cavity. This
proposal operates essentially through the dispersive coupling of each cavity
with the qutrit. The gate realization is quite simple because it requires
only a single-step operation. There is no need of applying a classical pulse
or performing a measurement. The gate operation time is independent of the
number of target qubits, thus it does not increase as the number of target
qubits increases. Moreover, because the third higher energy level of the
qutrit is not occupied during the gate operation, decoherence from the
qutrit is greatly suppressed. As an application of this hybrid
multi-target-qubit gate, we further discuss the generation of a hybrid
Greenberger-Horne-Zeilinger (GHZ) entangled state of SC qubits and cat-state
qubits. As an example, we numerically analyze the experimental feasibility of
generating a hybrid GHZ state of one SC qubit and three cat-state qubits within present circuit QED technology.
This proposal is quite general and can be extended to implement a hybrid controlled-NOT gate with one
matter qubit (of different type) simultaneously controlling multiple target
cat-state qubits in a wide range of physical systems, such as multiple
microwave or optical cavities coupled to a three-level natural or artificial
atom.
\end{abstract}

\maketitle
\date{\today }

\address{$^1$Department of Physics, Hangzhou Normal University, Hangzhou
311121, China }
\address{$^2$School of Physics, Nanjing University, Nanjing
210093, China}
\address{$^3$Quantum Information Research Center, Shangrao
Normal University, Shangrao 334001, China}

\begin{center}
\textbf{I. INTRODUCTION AND MOTIVATION}
\end{center}

Circuit quantum electrodynamics (QED), composed of microwave cavities and
superconducting (SC) qubits, has appeared as one of the most promising
candidates for quantum information processing (QIP) [1-7]. SC qubits (e.g.,
flux qubits, transmon qubits, Xmon qubits, fluxonium qubits, etc.) can be
flexibly and easily fabricated with microwave cavities, by using modern
integrated circuit technology. After more than 20 years of development, the
coherence time of SC qubits has leaped from the initial order of nanoseconds
to the order of hundreds of microseconds [8,9] or even over one millisecond
[10,11]. Due to their improved long coherence time and energy-level spacings
being rapidly adjustable [12-14], SC qubits are often used as the basic
processing units for solid-state QIP. So far, quantum gates with 2 SC qubits
[15-27], 3 SC qubits [28,29], and 4 SC qubits [29] have been experimentally
demonstrated. In addition, quantum entangled states of 2-6 SC qubits
[12,18,30,31], 10-12 SC qubits [19,32,33], and 18-20 SC qubits [34,35] have
been created in experiments.

On the other hand, one-dimensional (1D) microwave resonators of high quality
factor $Q\gtrsim 10^{6}$\ [36-41] and three-dimensional (3D) microwave
cavities with high quality factor $Q\gtrsim 3.5\times 10^{7}$\ [42-44] have
been reported in experiments. Owing to the experimental demonstration of
high-quality factors, microwave cavities or resonators contain microwave
photons with a lifetime comparable to that of SC qubits [45]. Quantum state
engineering and QIP using microwave fields and photons have recently drawn
much attention. In particular, there is an increasing interest in QIP with
cat-state qubits (i.e., qubits encoded with cat states) due to their
enhanced lifetime with quantum error correction (QEC) [46]. Various
proposals have been presented for implementing a universal set of single-
and two-qubit gates with cat-state qubits [47-52], implementing a
multi-qubit M\o lmer-S\o rensen \textit{entangling} gate using cat-state
qubits [53], and realizing a multi-target-qubit controlled-phase gate of
cat-state qubits [54]. Schemes have also been proposed for creating \textit{%
non-hybrid} Greenberger-Horne-Zeilinger (GHZ) entangled states of cat-state
qubits [55,53], preparing $W$-type entangled states of cat-state qubits
[56], and transferring entangled states of cat-state qubits [57]. In
addition, fault-tolerant quantum computation [58,59] and holonomic geometric
quantum control [60] of cat-state qubits have been studied. Moreover,
experiments have demonstrated a set of single-cat-state-qubit gates [61,62]
and produced entangled Bell states of two cat-state qubits [63].

The focus of this work is on hybrid quantum gates acting on different types
of qubits. Hybrid quantum gates have recently drawn considerable attention.
They play significant roles in connecting quantum information processors
with qubits of different encoding and have important applications in the
transmission of quantum states between a quantum processor and a quantum
memory. Over the past years, based on cavity or circuit QED, theoretical
proposals have been presented for implementing a hybrid two-qubit
controlled-phase or controlled-NOT gate with a charge qubit and an atomic
qubit [64], a charge qubit and a cat-state qubit [65], a photonic qubit and
a SC qubit [66], a photonic qubit and a cat-state qubit [67,68], and so on.
However, after a deep search of literature, we find that how to implement a
\textit{hybrid} multiqubit quantum gate with SC qubits and cat-state qubits
has not been investigated yet. The hybrid multiqubit gates of SC qubits and
cat-state qubits are of significance in realizing a large-scale QIP executed
in a compounded information processor, which is composed of a SC-qubit based
quantum processor and a cat-state-qubit based quantum processor. They are
also important in the transmission of quantum states between a SC-qubit
based quantum processor and a cat-state-qubit based quantum memory.
Recently, the architecture consisting of a SC processor and a quantum memory
has been shown to provide a significant interest [69].

\begin{figure}[tbp]
\begin{center}
\includegraphics[bb=179 197 536 455, width=8.5 cm, clip]{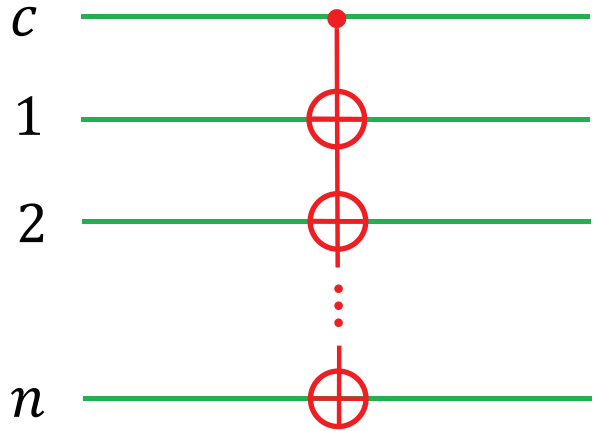} \vspace*{%
-0.08in}
\end{center}
\caption{(color online) Schematic circuit of a hybrid multi-target-qubit
gate. Here, $c$ represents the control SC qubit, while $1,2,...,n$ represent
the $n$ cat-state qubits. The two logic states of the SC qubit are
represented by the two lowest levels $|g\rangle$ and $|e\rangle$ of a
three-level SC qutrit, while the two logic states of each target cat-state
qubit are represented by the two quasi-orthogonal cat states $|0\rangle$ and
$|1\rangle$ (see equation (1) in the text). In the circuit, the symbol $%
\oplus $ represents a NOT gate on each target cat-state qubit. If the
control SC qubit is in the state $|e\rangle$, then the state at $\oplus $
for each target cat-state qubit is bit flipped as $\left\vert 0\right\rangle
\rightarrow \left\vert 1\right\rangle $ and $\left\vert 1\right\rangle
\rightarrow \left\vert 0\right\rangle $. However, when the control SC qubit
is in the state $|g\rangle$, the state at $\oplus $ for each target
cat-state qubit remains unchanged.}
\end{figure}

In the following, we will propose a method to \textit{directly} implement a
hybrid multi-target-qubit gate, i.e., a controlled-NOT gate with one SC
qubit (the control qubit) simultaneously controlling multiple target
cat-state qubits (Fig.~1). As is well known, a multi-target-qubit quantum
gate plays important roles in QIP. For instance, a multi-target-qubit
quantum gate has applications in entanglement preparation [70], quantum
cloning [71], error correction [72], and quantum algorithms [73].
Traditionally, a multi-target-qubit gate can in principle be constructed by
using single-qubit and two-qubit basic gates [74,75]. However, the number of
single-qubit and two-qubit gates, required to construct a multi-target-qubit
gate, increases substantially with the number of qubits. As a result, the
gate operation time will be quite long and decoherence significantly
degrades the gate fidelity. Therefore, it is worth finding effective ways to
\textit{directly} implement multi-target-qubit quantum gates. We should
point out that over the past years, many efficient methods have been
proposed to directly realize multi-target-qubit gates in various physical
systems [54,76-80]. However, it is noted that the previous works [54,76-80]
focus on the implementation of a \textit{non-hybrid} multi-target-qubit
gate. They are different from the present work which aims at implementing a
\textit{hybrid }multi-target-qubit gate\textit{\ }.

As an application of the proposed hybrid multi-target-qubit gate, we further
discuss the generation of hybrid GHZ entangled states of SC qubits and
cat-state qubits. We also numerically analyze the experimental feasibility
of preparing a hybrid GHZ entangled state of one SC qubit and three
cat-state qubits within present-day circuit QED. This proposal is quite
general and can be extended to implement a hybrid controlled-NOT gate with
one matter qubit (e.g., an atomic qubit, a quantum dot qubit, an NV center
qubit, or a SC qubit of different type) simultaneously controlling multiple
target cat-state qubits in a wide range of physical systems, such as
multiple microwave or optical cavities coupled to a three-level natural or
artificial atom (e.g., an atom, a quantum dot, an NV center, or a SC qutrit
of different type).

We stress that this work is different from Ref. [54] in the following
aspects. First, Ref.~[54] discussed how to implement a \textit{non-hybrid}
controlled-phase gate with one cat-state qubit simultaneously controlling
multiple target cat-state qubits, while the present work focuses on the
implementation of a \textit{hybrid} controlled-NOT gate with one SC qubit
simultaneously controlling multiple target cat-state qubits. The control
qubit in [54] is a cat-state qubit, but the control qubit in this work is a
SC qubit. Second, the gate in [54] was implemented by a pairwise
cavity-cavity interaction (see the effective Hamiltonian described by
Eq.~(7) in [54]); In contrast, the present gate is realized by an
interaction between each cavity and the coupler SC qutrit (see the effective
Hamiltonian given in Eq.~(6) below). Therefore, the focuses of the two works
and the physical mechanisms used in the two works are different.

We would like to point out that our work is different from Ref. [53]. First,
our work is for the implementation of a hybrid controlled-NOT gate with one
SC qubit simultaneously controlling $n$ target cat-state qubits, while Ref.
[53] is for the realization of a multi-qubit M\o lmer-S\o rensen \textit{%
entangling} gate using cat-state qubits. Our gate operates within a large $%
2^{n+1}$-dimensional Hilbert space, while the entangling gate in [53] is
manipulated only within a two-dimensional Hilbert space formed by two
product states of cat-state qubits. Second, our gate implementation does not
require the use of a classical pulse, while the gate realization in [53]
requires applying a classical pulse to each cavity in order to obtain a
two-photon drive on each cavity. Last, the effective Hamiltonian used for
the gate realization in our work is different from the effective Hamiltonian
applied for the gate implementation in [53] (see Eq.~(2) there).

This paper is organized as follows. In Sec. II, we briefly introduce a
hybrid controlled-NOT gate with one SC qubit simultaneously controlling
multiple target cat-state qubits. In Sec. III, we explicitly show how to
realize this hybrid multi-target-qubit gate. In Sec. IV, we show how to
generate \textit{hybrid} GHZ entangled states of SC qubits and cat-state
qubits by applying this gate. In Sec. V, we give a discussion on the
experimental feasibility of creating a hybrid GHZ state of one SC qubit and
three cat-state qubits by employing three 1D microwave cavities coupled to a
SC transmon qutrit. A concluding summary is given in Sec. VI.

\begin{center}
\textbf{II. HYBRID CONTROLLED-NOT GATE WITH A SC QUBIT SIMULTANEOUSLY
CONTROLLING MULTIPLE TARGET CAT-STATE QUBITS }
\end{center}

In this work, we will propose a method to realize a hybrid controlled-NOT
gate with one SC qubit simultaneously controlling $n$\ target cat-state
qubits ($1,2,...,n$) (Fig.~1). This multi-target-qubit gate is implemented
using $n$\ microwave cavities ($1,2,...,n$) coupled to a SC qutrit (Fig.
2a). In our proposal, the two logic states of the SC qubit (the control
qubit) are represented by the two lowest levels $\left\vert g\right\rangle $%
\ and $\left\vert e\right\rangle $\ of the SC qutrit, while the two logic
states of each target cat-state qubit are represented by two
quasi-orthogonal cat states of a microwave cavity. Namely, for cat-state
qubit $j$, the two logical states $|0\rangle _{j}$\ and $|1\rangle _{j}$\
are encoded with two cat states of cavity $j$, i.e.,
\begin{eqnarray}
|0\rangle _{j} &=&N_{\alpha }(|\alpha \rangle _{j}+|-\alpha \rangle _{j}),%
\text{ }  \notag \\
\text{\ }|1\rangle _{j} &=&N_{\alpha }(|i\alpha \rangle _{j}+|-i\alpha
\rangle _{j}),
\end{eqnarray}%
where $N_{\alpha }=1/\sqrt{2(1+e^{-2|\alpha |^{2}})}$ is the normalization
coefficient, $|\pm \alpha \rangle $\ and $|\pm i\alpha \rangle $\ are
coherent states with a real number $\alpha $. By taking $\alpha $\ large
enough, the two cat states $|0\rangle _{j}$\ and $|1\rangle _{j}$\ are
quasi-orthogonal to each other (e.g., for $\alpha =1.25,$\ one has $%
_{j}\left\langle 0\right\vert \left. 1\right\rangle _{j}=\cos \alpha
^{2}/\cosh \alpha ^{2}<10^{-3}$).

For $n$\ target qubits, there exist $2^{n}$\ computational basis states,
which form a set of complete orthogonal bases in a $2^{n}$-dimensional
Hilbert space of the $n$\ qubits. An $n$-target-qubit computational basis
state is denoted as $\left\vert l_{1}\right\rangle \left\vert
l_{2}\right\rangle ...\left\vert l_{n}\right\rangle $, where subscripts $%
1,2,...,n$ represent target qubits $1,2,...,n$ respectively, and $l_{j}\in
\{0,1\}$\ ($j=1,2,...,n$). In the present case that each target qubit is a
cat-state qubit, the two logic basis states $|0\rangle _{j}$\ and $|1\rangle
_{j}$\ of target qubit $j$ ($j=1,2,...,n$) are encoded as shown in Eq. (1).
The hybrid controlled-NOT gate, with one SC qubit simultaneously controlling
$n$\ target cat-state qubits, is characterized by the following state
transformation:%
\begin{eqnarray}
\left\vert g\right\rangle \left\vert l_{1}\right\rangle \left\vert
l_{2}\right\rangle ...\left\vert l_{n}\right\rangle &\rightarrow &\left\vert
g\right\rangle \left\vert l_{1}\right\rangle \left\vert l_{2}\right\rangle
...\left\vert l_{n}\right\rangle ,  \notag \\
\left\vert e\right\rangle \left\vert l_{1}\right\rangle \left\vert
l_{2}\right\rangle ...\left\vert l_{n}\right\rangle &\rightarrow &\left\vert
e\right\rangle \left\vert \overline{l}_{1}\right\rangle \left\vert \overline{%
l}_{2}\right\rangle ...\left\vert \overline{l}_{n}\right\rangle ,
\end{eqnarray}%
where $\overline{l}_{j}=1-l_{j}$\ $\in \{0,1\}$\ ($j=1,2,...,n$). This
transformation (2) implies that only when the control SC qubit (the first
qubit) is in the state $\left\vert e\right\rangle $, a bit flip happens to
the state of each of the $n$\ target cat-state qubits, i.e., $\left\vert
0\right\rangle \rightarrow \left\vert 1\right\rangle $\ and $\left\vert
1\right\rangle \rightarrow \left\vert 0\right\rangle $\ for each target
cat-state qubit, while nothing happens to the state of each target cat-state
qubit when the control SC qubit is in the state $\left\vert g\right\rangle .$
In the next section, we will show how to implement this hybrid
multi-target-qubit gate.

\begin{figure}[tbp]
\begin{center}
\includegraphics[bb=62 4 850 484, width=11.5 cm, clip]{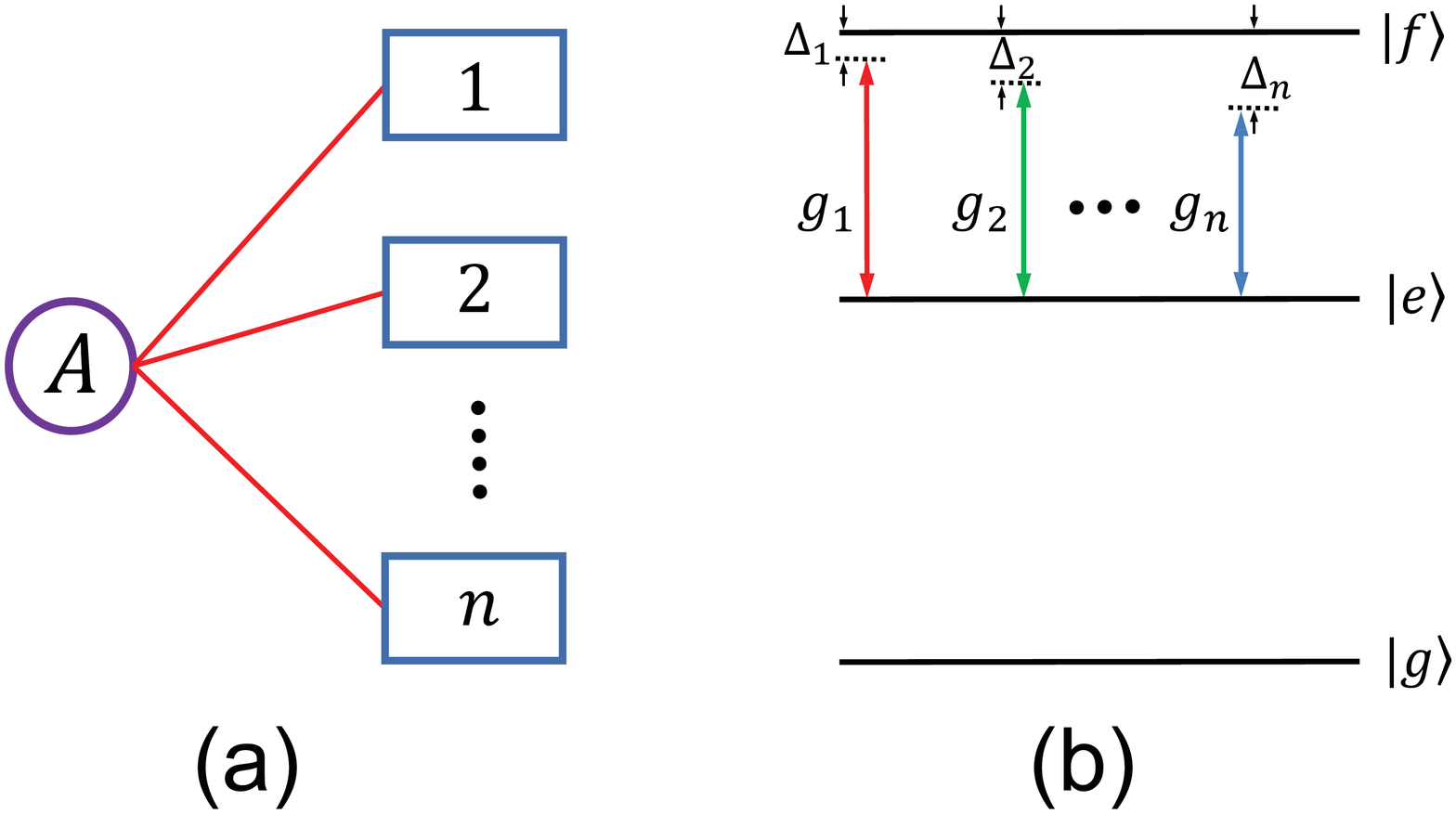} \vspace*{%
-0.08in}
\end{center}
\caption{(color online) (a) Schematic circuit of $n$ microwave cavities
coupled to a SC qutrit. Each square represents a one-dimensional (1D) or
three-dimensional (3D) microwave cavity. The circle $A$ represents the SC
qutrit, which is inductively or capacitively coupled to each cavity. (b)
Illustration of $n$ cavities $(1,2,...,n)$ dispersively coupled to the $%
\left\vert e\right\rangle \leftrightarrow \left\vert f\right\rangle $
transition of the qutrit. In (b), the level spacing between the upper two
levels is smaller than that between the two lowest levels, which applies to
a SC phase, transmon, Xmon, or fluxonium qutrit. Alternatively, the level
spacing between the upper two levels can be larger than that between the two
lowest levels, which applies to a SC charge, flux, or fluxonium qutrit, etc..
}
\end{figure}

\begin{center}
\textbf{III. IMPLEMENTATION OF THE HYBRID MULTI-TARGET-QUBIT CONTROLLED-NOT
GATE }
\end{center}

The physical system consists of $n$\ microwave cavities ($1,2,...,n$)
coupled to a SC qutrit (Fig.~2a). We define the three levels of the qutrit
as $|g\rangle $, $|e\rangle $\ and $|f\rangle $\ (Fig.~2b). Suppose that
cavity $j$\ ($j=1,2,...,n$) is dispersively coupled to the $|e\rangle
\leftrightarrow |f\rangle $ transition of the qutrit with coupling constant $%
g_{j}$ and detuning $\Delta _{j}$ (Fig. 2b). In addition, assume that the
couplings of each cavity with the $|g\rangle \leftrightarrow |e\rangle $ and
$|g\rangle \leftrightarrow |f\rangle $ transitions of the qutrit are not
considered in our theoretical model. Note that the coupling and decoupling
between the SC qutrit and each cavity can in principle be achieved by prior
adjustment of the level spacings of the coupler qutrit or/and prior
adjustment of the cavity frequency. For a SC qutrit, the level spacings can
be rapidly (within 1-3 ns) adjusted by changing external control parameters
(e.g., magnetic flux applied to the superconducting loop of a SC qutrit
[12-14,18,81]). Moreover, the frequency of a microwave cavity or resonator
can be quickly tunned within a few nanoseconds [82,83].

When the above assumptions are applied, the Hamiltonian of the entire
system, in the interaction picture and after making the rotating-wave
approximation (RWA), is given by (in units of $\hbar =1$)
\begin{equation}
H_{\mathrm{I}}=\sum\limits_{j=1}^{n}g_{j}(e^{i\Delta _{j}t}\hat{a}%
_{j}|f\rangle \langle e|+\text{h.c.}),
\end{equation}%
where $\hat{a}_{j}$\ is the photon annihilation operator of cavity $j,$\ and
$\Delta _{j}=\omega _{fe}-\omega _{c_{j}}$. Here, $\omega _{fe}$ is the $%
|e\rangle \leftrightarrow |f\rangle $\ transition frequency of the qutrit
while $\omega _{c_{j}}$ is the frequency of cavity $j$\ ($j=1,2,...,n$).

In the case of $\left\vert \Delta _{j}\right\vert \gg g_{j}$\
(large-detuning condition)$,$\ the energy exchange does not happen between
the qutrit and cavity $j$\ ($j=1,2,...,n$). In addition, when
\begin{equation}
\frac{\left\vert \Delta _{j}-\Delta _{k}\right\vert }{\left\vert \Delta
_{j}^{-1}\right\vert +\left\vert \Delta _{k}^{-1}\right\vert }\gg g_{j}g_{k},
\end{equation}%
(where $j,k\in \{1,2,...,n\},$\ $j\neq k$), the qutrit does not induce the
interaction between any two of the cavities. Under these considerations, it
is straightforward to show that the Hamiltonian (3) becomes [84-86]
\begin{equation}
H_{\mathrm{e}}=-\sum\limits_{j=1}^{n}\lambda _{j}(\hat{a}_{j}^{+}\hat{a}%
_{j}|e\rangle \langle e|-\hat{a}_{j}\hat{a}_{j}^{+}|f\rangle \langle f|),
\end{equation}%
where $\lambda _{j}=g_{j}^{2}/\Delta _{j}$. In Eq. (5), the first (second)\
term describes the photon-number dependent Stark shift of the energy level $%
|e\rangle $\ ($|f\rangle $), induced by the cavities. In the case when the
level $|f\rangle $\ is not occupied, the Hamiltonian (5) further reduces to
\begin{equation}
H_{\mathrm{e}}=-\sum\limits_{j=1}^{n}\lambda _{j}\hat{a}_{j}^{+}\hat{a}%
_{j}|e\rangle \langle e|.
\end{equation}%
We set $\lambda _{j}=\lambda $\ ($j=1,2,...,n$), i.e., $\lambda _{1}=\lambda
_{2}=...=\lambda _{n}$ which turns out into
\begin{equation}
g_{1}^{2}/\Delta _{1}=g_{2}^{2}/\Delta _{2}=...=g_{n}^{2}/\Delta _{n}.
\end{equation}%
Because of $\Delta _{j}=\omega _{fe}-\omega _{c_{j}},$\ this equality (7)
can be met by carefully selecting the detuning $\Delta _{j}$\ via tuning the
frequency $\omega _{c_{j}}$\ of cavity $j$\ ($j=1,2,...,n$).

For the Hamiltonian $H_{\mathrm{e}}$\ given in Eq. (6), the unitary operator
$U=e^{-iH_{\mathrm{e}}t}$\ can be written as
\begin{equation}
U=\prod\limits_{j=1}^{n}U_{j},
\end{equation}%
where $U_{j}$\ is a unitary operator acting on cavity $j$\ and the qutrit,
which is given by
\begin{equation}
U_{j}=\exp \left( i\lambda \hat{a}_{j}^{+}\hat{a}_{j}|e\rangle \langle
e|t\right) .
\end{equation}

Based on Eq. (1) and Eq. (9), it is straightforward to show that the unitary
operator $U_{j}$\ results in the following state transformation
\begin{align}
U_{j}|g\rangle |0\rangle _{j}& =|g\rangle |0\rangle _{j},  \notag \\
U_{j}|g\rangle |1\rangle _{j}& =|g\rangle |1\rangle _{j},  \notag \\
U_{j}|e\rangle |0\rangle _{j}& =|e\rangle N_{\alpha }(|e^{i\lambda t}\alpha
\rangle _{j}+|-e^{i\lambda t}\alpha \rangle _{j}),  \notag \\
U_{j}|e\rangle |1\rangle _{j}& =|e\rangle N_{\alpha }(|e^{i\lambda t}i\alpha
\rangle _{j}+|-e^{i\lambda t}i\alpha \rangle _{j}).
\end{align}%
For $\lambda t=\pi /2,$\ we have
\begin{eqnarray}
N_{\alpha }\left( |e^{i\lambda t}\alpha \rangle _{j}+|-e^{i\lambda t}\alpha
\rangle _{j}\right) &=&N_{\alpha }\left( |i\alpha \rangle _{j}+|-i\alpha
\rangle _{j}\right) =|1\rangle _{j},  \notag \\
N_{\alpha }\left( |e^{i\lambda t}i\alpha \rangle _{j}+|-e^{i\lambda
t}i\alpha \rangle _{j}\right) &=&N_{\alpha }\left( |-\alpha \rangle
_{j}+|\alpha \rangle _{j}\right) =|0\rangle _{j}.
\end{eqnarray}%
\ Based on Eq. (11), the state transformation (10) thus becomes
\begin{align}
U_{j}|g\rangle |0\rangle _{j}& =|g\rangle |0\rangle _{j},  \notag \\
U_{j}|g\rangle |1\rangle _{j}& =|g\rangle |1\rangle _{j},  \notag \\
U_{j}|e\rangle |0\rangle _{j}& =|e\rangle |1\rangle _{j},  \notag \\
U_{j}|e\rangle |1\rangle _{j}& =|e\rangle |0\rangle _{j},
\end{align}%
which implies that when the SC qubit is initially in the state $\left\vert
e\right\rangle ,$\ the state $|0\rangle _{j}$\ ($|1\rangle _{j}$) of
cat-state qubit $j$\ flips to the state $|1\rangle _{j}$\ ($|0\rangle _{j}$%
); otherwise, nothing happens to the state of cat-state qubit $j$\ when the
SC qubit is initially in the state $\left\vert g\right\rangle $.

According to Eqs. (8) and (12), it is easy to obtain the following state
transformation%
\begin{eqnarray}
\prod\limits_{j=1}^{n}U_{j}\left\vert g\right\rangle \left\vert
l_{1}\right\rangle \left\vert l_{2}\right\rangle ...\left\vert
l_{n}\right\rangle &=&\left\vert g\right\rangle \left\vert
l_{1}\right\rangle \left\vert l_{2}\right\rangle ...\left\vert
l_{n}\right\rangle ,  \notag \\
\prod\limits_{j=1}^{n}U_{j}\left\vert e\right\rangle \left\vert
l_{1}\right\rangle \left\vert l_{2}\right\rangle ...\left\vert
l_{n}\right\rangle &=&\left\vert e\right\rangle \left\vert \overline{l}%
_{1}\right\rangle \left\vert \overline{l}_{2}\right\rangle ...\left\vert
\overline{l}_{n}\right\rangle ,
\end{eqnarray}%
where $l_{j},\overline{l}_{j}\in \{0,1\},$\ and $\overline{l}_{j}=1-l_{j}$\ (%
$j=1,2,...,n$).

The result (13) shows that when the control SC qubit is in the state $%
\left\vert g\right\rangle $, nothing happens to the state of each of the $n$%
\ target cat-state qubits ($1,2,...,n$); however, when the control SC qubit
is in the state $\left\vert e\right\rangle $, a bit flip from $\left\vert
0\right\rangle $\ to $\left\vert 1\right\rangle $\ (from $\left\vert
1\right\rangle $\ to $\left\vert 0\right\rangle $) happens to the state $%
\left\vert 0\right\rangle $\ ($\left\vert 1\right\rangle $) of each of the $%
n $\ target cat-state qubits ($1,2,...,n$). Hence, a hybrid
multi-target-qubit controlled-NOT gate, described by Eq. (2), is implemented
with a SC qubit (the control qubit) and $n$\ target cat-state qubits ($%
1,2,...,n$), after the above operation.

As shown above, this hybrid multi-target-qubit gate is realized through a
single unitary operator $U.$\ Note that we derived the unitary operator $U$\
by starting with the original Hamiltonian (3). Therefore, the gate
realization only requires a single-step operation described by $U$. The
third higher energy level $\left\vert f\right\rangle $\ of the qutrit is not
occupied during the gate operation. Thus, decoherence from this level of the
qutrit is greatly suppressed. In addition, neither applying a classical
pulse to the qutrit nor performing a measurement on the state of the qutrit/
cavities is needed for the gate realization.

\begin{center}
\textbf{IV. GENERATION OF HYBRID GHZ ENTANGLED STATES}
\end{center}

Hybrid entangled states play a key role in QIP and quantum technology. For
example, they help to answer fundamental questions, such as the boundary
between quantum and classical domains, and the so-called Schr\"{o}dinger's
paradox [87], where microscopic quantum systems and macroscopic classical
systems are entangled with each other. Moreover, hybrid entangled states can
be used as an important quantum channel and intermediate resource for
quantum technology, covering the transmission, operation, and storage of
quantum information between different formats and encodings [88-90]. On the
other hand, GHZ entangled states are not only of great interest for
fundamental tests of quantum mechanics [91], but also have applications in
QIP [92], quantum communications[93], error-correction protocols [94],
quantum metrology [95], and high-precision spectroscopy [96]. In this
section, we discuss how to create a hybrid GHZ entangled state of SC qubits
and cat-state qubits by applying the gate above.

\begin{center}
\textbf{A. Generation of hybrid GHZ entangled states with }$one$\textbf{\ SC
qubit and }$n$\textbf{\ cat-state qubits}
\end{center}

Let us return to the physical system depicted in Fig. 2a. Assume that the SC
qutrit is initially in the state $\left\vert +\right\rangle =\left(
|g\rangle +|e\rangle \right) /\sqrt{2},$\ which can be readily prepared by
applying a classical pulse (resonant with the $\left\vert g\right\rangle
\leftrightarrow \left\vert e\right\rangle $\ transition) to the qutrit in
the ground state $\left\vert g\right\rangle $\ [97]. In addition, assume
that each cavity is initially in the cat state $|0\rangle =N_{\alpha
}(|\alpha \rangle +|-\alpha \rangle ).$\ Experimentally, the cat state here
has been produced in circuit QED [62,98-101]. The initial state of the whole
system is thus given by
\begin{equation}
\left\vert \psi \left( 0\right) \right\rangle =\frac{1}{\sqrt{2}}\left(
|g\rangle +|e\rangle \right) |0\rangle _{1}|0\rangle _{2}...|0\rangle _{n}.
\end{equation}%
Now apply the hybrid controlled-NOT gate (13) [i.e., the gate in Eq. (2)] on
the qutrit and the $n$\ cat-state qubits. According to Eq. (13), it is easy
to see that the state $\left\vert \psi \left( 0\right) \right\rangle $\
becomes
\begin{equation}
|\text{GHZ}\rangle _{\mathrm{hyb}}=\frac{1}{\sqrt{2}}\left( |g\rangle
|0\rangle _{1}|0\rangle _{2}...|0\rangle _{n}+|e\rangle |1\rangle
_{1}|1\rangle _{2}...|1\rangle _{n}\right) ,
\end{equation}%
which shows that one SC qubit and $n$ cat-state qubits are prepared in a
hybrid GHZ entangled state.

\begin{figure}[tbp]
\begin{center}
\includegraphics[bb=105 160 800 452, width=12.0 cm, clip]{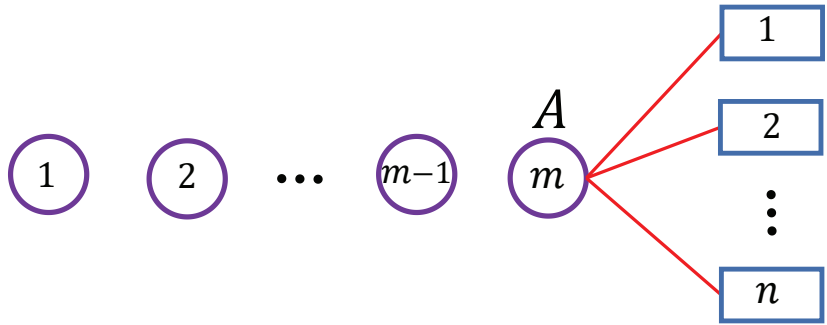} \vspace*{%
-0.08in}
\end{center}
\caption{(color online) Schematic of setup of $m$ SC qubits ($1,2,...,m$)
and $n$ microwave cavities ($1,2,...,n$). Here, the $m$th SC qubit is formed
by the two lowest levels of a SC qutrit $A$ (see Fig.~2b), which is coupled
to the $n$ cavities. Note that the remaining $m-1$ SC qubits ($1,2,...,m-1$)
are decoupled from the $n$ cavities. The $m-1$ circles on the left represent
the $m-1$ SC qubits ($1,2,...,m-1$), while the rightmost circle represents
the $m$th SC qubit or the coupler qutrit $A$. Eeach square represents a 1D
or 3D microwave cavity.}
\label{fig:3}
\end{figure}

\begin{center}
\textbf{B. Generation of hybrid GHZ entangled states with }$m$\textbf{\ SC
qubits and }$n$\textbf{\ cat-state qubits}
\end{center}

Let us consider $m$ SC qubits ($1,2,...,m)$ and $n$ microwave cavities ($%
1,2,...,n$). The $n$ cavities are coupled to a SC qutrit $A$ (Fig. 3). Note
that the $m$th SC qubit here are formed by the two lowest levels of the
coupler SC qutrit $A$. Assume that each cavity is initially in the cat state
$|0\rangle =N_{\alpha }(|\alpha \rangle +|-\alpha \rangle ).$\ In addition,
assume that the $m$ SC qubits ($1,2,...,m)$ are initially in the GHZ state $%
\left\vert GHZ\right\rangle =\left( |g\rangle _{1}|g\rangle _{2}...|g\rangle
_{m}+|e\rangle _{1}|e\rangle _{2}...|e\rangle _{m}\right) /\sqrt{2}.$\ In
the past years, theoretical schemes for preparing multi-SC-qubit GHZ
entangled states have been proposed [102-107], and the experimental
production of GHZ entangled states with up to 18 SC qubits has been reported
[12,18,19,30-34]. The initial state of the whole system is thus given by
\begin{equation}
\left\vert \psi \left( 0\right) \right\rangle ^{\prime }=\frac{1}{\sqrt{2}}%
\left( |g\rangle _{1}|g\rangle _{2}...|g\rangle _{m}+|e\rangle _{1}|e\rangle
_{2}...|e\rangle _{m}\right) |0\rangle _{1}|0\rangle _{2}...|0\rangle _{n}.
\end{equation}%
The first $m-1$ SC qubits ($1,2,...,m-1)$ are not coupled to the $n$
cavities, while the $m$th SC qubit (i.e., the coupler SC qutrit $A$) is
dispersively coupled to the $n$ cavities such that a hybrid controlled-NOT
gate (13) is performed on the $m$th SC qubit and the $n$\ cat-state qubits.
The details on the implementation of the gate (13) are presented in the
previous section III. According to Eq. (13), it is easy to see that after
the gate operation, the state $\left\vert \psi \left( 0\right) \right\rangle
^{\prime }$\ becomes
\begin{equation}
|\text{GHZ}\rangle _{\mathrm{hyb}}=\frac{1}{\sqrt{2}}\left( g\rangle
_{1}|g\rangle _{2}...|g\rangle _{m-1}|g\rangle _{m}|0\rangle _{1}|0\rangle
_{2}...|0\rangle _{n}|g\rangle +|e\rangle _{1}|e\rangle _{2}...|e\rangle
_{m-1}|e\rangle _{m}|1\rangle _{1}|1\rangle _{2}...|1\rangle _{n}\right) ,
\end{equation}%
which shows that the $m$ SC qubits ($1,2,...,m)$ and the $n$ cat-state
qubits ($1,2,...,n)$ are prepared in a hybrid GHZ entangled state.

From the descriptions given above, one can see that the hybrid GHZ states of
SC qubits and cat-state qubits can be straightforwardly created by applying
the hybrid controlled-NOT gate (2) or (13). Given that the initial state $%
\left\vert \psi \left( 0\right) \right\rangle $\ or $\left\vert \psi \left(
0\right) \right\rangle ^{\prime }$ of the system is ready, the operation
time for the preparation of the hybrid GHZ state (15) or (17) is equal to
that for the implementation of the gate (13), i.e., $t=\pi /\left( 2\lambda
\right) $. Since the GHZ states here are created based on the gate (13), the
Hamiltonians used for the GHZ state production are the same as those used
for the implementation of the gate (13).

Before ending this section, it should be mentioned that both Refs. [53,55]
have proposed how to create \textit{non-hybrid} GHZ entangled states of
multiple cat-state qubits. However, it is noted that how to prepare a
\textit{hybrid} GHZ entangled state with SC qubits and cat-state qubits was
not studied in both of [53,55].

\begin{figure}[tbp]
\begin{center}
\includegraphics[bb=311 141 552 376, width=7.0 cm, clip]{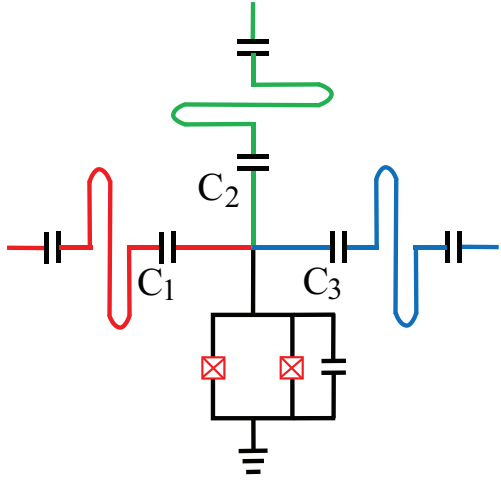} \vspace*{%
-0.08in}
\end{center}
\caption{(color online) Diagram of three 1D microwave cavities capacitively
coupled to a SC transmon qutrit. Each cavity here is a one-dimensional
transmission line resonator. The transmon qutrit consists of two Josephson
junctions and a capacitor.}
\label{fig:4}
\end{figure}

\begin{center}
\textbf{V. POSSIBLE EXPERIMENTAL IMPLEMENTATION}
\end{center}

In this section, as an example, we investigate the experimental feasibility
for creating a hybrid GHZ state with one\ SC qubit and three\ cat-state
qubits [i.e., the GHZ state (15) with $n=3$], by using three 1D microwave
cavities coupled to a SC transmon qutrit (Fig.~4). As shown in subsection IV
A, the GHZ state (15) was prepared by applying the hybrid multi-target-qubit
controlled-NOT gate (13) [i.e., the gate in Eq. (2)]. In this sense, as long
as the initial state (14) can be well prepared, the operational fidelity for
the preparation of the GHZ state (15) mainly depends on the performance of
the hybrid gate (13) applied on the SC qubit and the three cat-state qubits.

\begin{center}
\textbf{A. Full Hamiltonian}
\end{center}

In the preceding section III, the hybrid gate (13) was realized based on the
effective Hamiltonian (6). Note that this Hamiltonian was derived from the
original Hamiltonian (3), which only contains the coupling between each
cavity and the $|e\rangle \leftrightarrow |f\rangle $\ transition of the SC
qutrit. In a realistic situation, there exists the unwanted coupling between
each cavity and the $|g\rangle \leftrightarrow |f\rangle $\ transition of
the SC qutrit as well as the unwanted coupling between each cavity and the $%
|g\rangle \leftrightarrow |e\rangle $\ transition of the SC qutrit. In
addition, there exists the unwanted inter-cavity crosstalk between the three
cavities.

\begin{figure}[tbp]
\begin{center}
\includegraphics[bb=211 91 765 463, width=9.0 cm, clip]{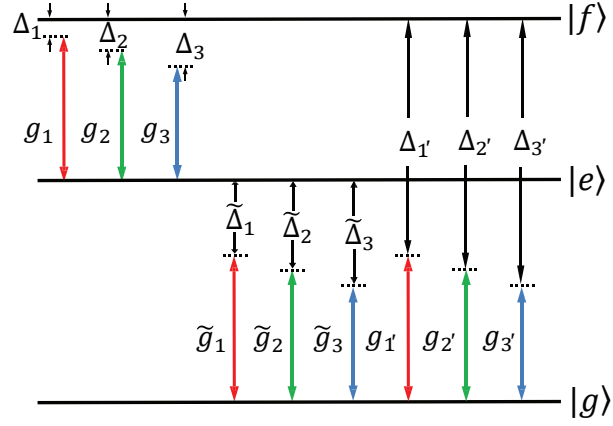} \vspace*{%
-0.08in}
\end{center}
\caption{(color online) Illustration of three cavities (1,2,3) coupled to
the $\left\vert e\right\rangle \leftrightarrow \left\vert f\right\rangle $
transition of the qutrit, as well as the unwanted couplings of the three
cavities with the $\left\vert g\right\rangle \leftrightarrow \left\vert
f\right\rangle $ transition and the $\left\vert g\right\rangle
\leftrightarrow \left\vert e\right\rangle $ transition of the qutrit.}
\label{fig:5}
\end{figure}

By taking the unwanted couplings and the unwanted inter-cavity crosstalk
into account, the Hamiltonian (3), with $n=3$\ for the present case, is
modified as
\begin{eqnarray}
H_{\mathrm{I}}^{\prime } &=&\sum\limits_{j=1}^{3}g_{j}(e^{i\Delta _{j}t}\hat{%
a}_{j}|f\rangle \langle e|+\text{h.c.})  \notag \\
&&+\sum\limits_{j=1}^{3}g_{j}^{\prime }(e^{i\Delta _{j}^{\prime }t}\hat{a}%
_{j}|f\rangle \langle g|+\text{h.c.})+\sum\limits_{j=1}^{3}\widetilde{g}%
_{j}(e^{i\widetilde{\Delta }_{j}t}\hat{a}_{j}|e\rangle \langle g|+\text{h.c.}%
)  \notag \\
&&+\left( g_{12}e^{i\Delta _{12}t}\hat{a}_{1}^{+}\hat{a}_{2}+\text{h.c.}%
\right) +\left( g_{13}e^{i\Delta _{13}t}\hat{a}_{1}^{+}\hat{a}_{3}+\text{h.c.%
}\right) +\left( g_{23}e^{i\Delta _{23^{\prime }}t}\hat{a}_{2}^{+}\hat{a}%
_{3}+\text{h.c.}\right) ,
\end{eqnarray}%
where the term in line one represents the required coupling of cavity $j$\
with $|e\rangle \leftrightarrow |f\rangle $\ transition of the SC qutrit
(Fig.~5), the first term in line two represents the unwanted coupling of
cavity $j$\ with $|g\rangle \leftrightarrow |f\rangle $\ transition of the
SC qutrit with the coupling constant $g_{j}^{\prime }$\ and the detuning $%
\Delta _{j}^{\prime }=\omega _{fg}-\omega _{c_{j}}$\ (Fig.~5), the second
term in line two represents the unwanted coupling of cavity $j$\ with $%
|g\rangle \leftrightarrow |e\rangle $\ transition of the SC qutrit with the
coupling constant $\widetilde{g}_{j}$\ and the detuning $\widetilde{\Delta }%
_{j}=\omega _{eg}-\omega _{c_{j}}$\ (Fig. 5), while the terms in the last
line represent the unwanted inter-cavity crosstalk among the three cavities,
$g_{kl}$\ is the crosstalk strength between the two cavities $k$\ and $l$\
with the frequency difference $\Delta _{kl}=\omega _{c_{k}}-\omega _{c_{l}}$%
\ $\left( k,l\in \{1,2,3\};k\neq l\right) $. Here, $\omega _{fg}$\ ($\omega
_{eg}$) is the $|g\rangle \leftrightarrow |f\rangle $\ ($|g\rangle
\leftrightarrow |e\rangle $)\ transition frequency of the qutrit, while $%
\omega _{c_{k}}$\ ($\omega _{c_{l}}$) is the frequency of cavity $k$\ ($l$).

\begin{center}
\textbf{B. Numerical results}
\end{center}

The dynamics of the lossy system, with finite qutrit relaxation, dephasing
and photon lifetime being included, is determined by%
\begin{eqnarray}
\frac{d\rho }{dt} &=&-i\left[ H_{\mathrm{I}}^{\prime },\rho \right]
+\sum_{j=1}^{3}\kappa _{j}\mathcal{L}\left[ \hat{a}_{j}\right]  \notag \\
&&+\gamma _{eg}\mathcal{L}\left[ \sigma _{eg}^{-}\right] +\gamma _{fe}%
\mathcal{L}\left[ \sigma _{fe}^{-}\right] +\gamma _{fg}\mathcal{L}\left[
\sigma _{fg}^{-}\right]  \notag \\
&&+\gamma _{e,\varphi }\left( \sigma _{ee}\rho \sigma _{ee}-\sigma _{ee}\rho
/2-\rho \sigma _{ee}/2\right)  \notag \\
&&+\gamma _{f,\varphi }\left( \sigma _{ff}\rho \sigma _{ff}-\sigma _{ff}\rho
/2-\rho \sigma _{ff}/2\right) ,
\end{eqnarray}%
where $\sigma _{eg}^{-}=\left\vert g\right\rangle \left\langle e\right\vert
, $\ $\sigma _{fe}^{-}=\left\vert e\right\rangle \left\langle f\right\vert ,$%
\ $\sigma _{fg}^{-}=\left\vert g\right\rangle \left\langle f\right\vert ,$\ $%
\sigma _{ee}=\left\vert e\right\rangle \left\langle e\right\vert $, $\sigma
_{ff}=\left\vert f\right\rangle \left\langle f\right\vert ,$\ $\mathcal{L}%
\left[ \Lambda \right] =\Lambda \rho \Lambda ^{+}-\Lambda ^{+}\Lambda \rho
/2-\rho \Lambda ^{+}\Lambda /2$\ (with $\Lambda =\hat{a}_{j},\sigma
_{eg}^{-},\sigma _{fe}^{-},\sigma _{fg}^{-})$,\ $\kappa _{j}$\ is the decay
rate of cavity $j$ ($j=1,2,3$),\ $\gamma _{eg}$ is the energy relaxation
rate of the level $\left\vert e\right\rangle $\ for the decay path $%
\left\vert e\right\rangle \rightarrow \left\vert g\right\rangle $\ of the
qutrit, $\gamma _{fe}$\ ($\gamma _{fg}$) is the relaxation rate of the level
$\left\vert f\right\rangle $\ for the decay path $\left\vert f\right\rangle
\rightarrow \left\vert e\right\rangle $\ ($\left\vert f\right\rangle
\rightarrow \left\vert g\right\rangle $) of the qutrit; $\gamma _{e,\varphi
} $\ ($\gamma _{f,\varphi }$) is the dephasing rate of the level $\left\vert
e\right\rangle $\ ($\left\vert f\right\rangle $) of the qutrit.

The operational fidelity is given by $F=\sqrt{\langle \psi _{\mathrm{id}%
}|\rho |\psi _{\mathrm{id}}\rangle },$\ where $|\psi _{\mathrm{id}}\rangle $%
\ is the ideal output state of Eq.~(15) with $n=3$\ (which is obtained
without considering the system dissipation, the inter-cavity crosstalk and
the unwanted couplings); while $\rho $\ is the density operator of the
system when the operation is performed in a realistic situation.\

\begin{table}[tbp]
\centering
\begin{tabular*}{12.2cm}{lll}
\hline\hline
\ \ $\omega_{eg}/2\pi=4.0 \mathrm{\,GHz}$ & \ \ \ \ $\omega_{fe}/2\pi=3.3
\mathrm{\,GHz}$ & \,\ \ \ $\omega_{fg}/2\pi=7.3 \mathrm{\,GHz}$ \\[1.5ex]
\ \ \,\!$\omega_{c_{1}}/2\pi=3.24 \mathrm{\,GHz}$ & \ \ \,\ \ $%
\omega_{c_{2}}/2\pi=3.21 \mathrm{\,GHz}$ & \ \ \ \ $\omega_{c_{3}}/2\pi= 3.18
\mathrm{\,GHz}$ \\[1.5ex]
\ \ \ \!\!$\triangle_{1}/2\pi= 60 \mathrm{\,MHz}$ & \quad \ \ $%
\triangle_{2}/2\pi= 90 \mathrm{\,MHz}$ & \ \,\ \ \ \!$\triangle_{3}/2\pi= 120
\mathrm{\,MHz}$ \\[1.5ex]
\ \ \ \!\!\!$\triangle_{1^{^{\prime }}}/2\pi= 4.06 \mathrm{\,GHz}$ & \ \ \ \ $%
\triangle_{2^{^{\prime }}}/2\pi= 4.09 \mathrm{\,GHz}$ & \ \,\,\ \ \!\!$%
\triangle_{3^{^{\prime }}}/2\pi= 4.12 \mathrm{\,GHz}$ \\[1.5ex]
\ \ \ \!$\widetilde{\triangle}_{1}/2\pi= 760 \mathrm{\,MHz}$ & \ \ \ \ \,\,\,\!$%
\widetilde{\triangle}_{2}/2\pi= 790 \mathrm{\,MHz}$ & \ \ \ \ $\widetilde{%
\triangle}_{3}/2\pi= 820 \mathrm{\,MHz}$ \\[1.5ex]
\!\!\!\quad $\triangle_{12}/2\pi= 30 \mathrm{\,MHz}$ & \ \;\,\ \ \!$%
\triangle_{23}/2\pi= 30 \mathrm{\,MHz}$ & \;\;\,\, \!\!\,$\triangle_{13}/2\pi=
60 \mathrm{\,MHz}$ \\[1.5ex]
\!\!\!\!\!\!\!\quad \ \ \,\,\, $g_{1}/2\pi= 4.5 \mathrm{\,MHz}$ & \ \ \ \ \
\,\,$g_{2}/2\pi= 5.51 \mathrm{\,MHz}$ & \quad \ \ \ \!$g_{3}/2\pi= 6.36
\mathrm{\,MHz}$ \\ \hline\hline
\end{tabular*}
\caption{For the definitions of the parameters, please refer to the text.
Except for $g_{2}$\ and $g_{3},$ all parameters listed are used in the
numberical simulations for plotting Figs. 6 and 7. The coupling constants $%
g_{2}$\ and $g_{3}$ listed are used for Fig. 6, which are calculated
according to Eq.~(7). While, the coupling constants $g_{2}$\ and $g_{3}$
used for Fig. 7 are not shown, which are dependent of $c$ [see Eq.~(21)] and
calculated according to Eq.~(21), by using the values of $g_{1},\Delta
_{1},\Delta _{2}, $ and $\Delta _{3}$\ listed in the table.}
\end{table}

As a concrete example, let us consider the parameters listed in Table 1,
which are used in our numerical simulations. For a transmon qutrit, the
level spacing anharmonicity 100-720 MHz was reported in experiments [108].
For a transmon qutrit [109], one has $\widetilde{g}_{j}=g_{j}/\sqrt{2}%
(j=1,2,,3)$. Since the $|g\rangle \leftrightarrow |f\rangle $ transition for
a transmon qutrit is forbidden or weak [109], we choose $g_{j}^{\prime
}=0.01g_{j}$ $(j=1,2,,3).$ For the coupling constants listed in Table 1, the
maximal value $g_{\max }=\max \{g_{1},g_{2},g_{3}\}$\ is $2\pi \times 6.36$\
MHz, which is readily available because a coupling strength $\sim 2\pi
\times 360$\ MHz has been reported for a transmon qutrit coupled to a 1D
microwave cavity [110].

Other parameters used in the numerical simulations are: (i) $\gamma
_{eg}^{-1}=60$\ $\mu $s, $\gamma _{fg}^{-1}=150$\ $\mu $s [111], $\gamma
_{fe}^{-1}=30$\ $\mu $s, $\gamma _{\phi e}^{-1}=\gamma _{\phi f}^{-1}=20$\ $%
\mu $s, (ii) $\kappa _{1}=\kappa _{2}=\kappa _{3}=\kappa $, (iii) $%
g_{12}=g_{23}=g_{13}=g_{cr}$,\ (iv) $g_{cr}=0.01g_{\max }$, and (v) $\alpha
=1.25.$\ Here, we consider a rather conservative case for decoherence time
of the transmon qutrit because energy relaxation time with a range from 65 $%
\mu $s to 0.5 ms and dephasing time from 25 $\mu $s to 75 $\mu $s have been
experimentally demonstrated for a superconducting transmon device [8,9,112,113].
In addition, for each cavity, a cat state with $\alpha =1.25$ can be created
in experiments because the circuit-QED experimental preparation of a cat
state with the amplitude $\left\vert \alpha \right\vert \leq 5.27$ has been
reported [62,98-101]. The choice of $g_{cr}=0.01g_{\max }$ is obtainable in
experiments by a prior design of the sample with appropriate capacitances $%
C_{1}$, $C_{2}$, and $C_{3}$\ depicted in Fig.~4 [114].

In a realistic situation, the initial state of Eq.~(14) may not be prepared
perfectly. Therefore, we consider a non-ideal initial state of the system%
\begin{eqnarray}
\left\vert \psi \left( 0\right) \right\rangle _{\mathrm{non-ideal}} &=&%
\widetilde{N}_{\alpha }^{3}\left( \sqrt{1+\delta }|\alpha \rangle _{1}+\sqrt{%
1-\delta }|-\alpha \rangle _{1}\right) \left( \sqrt{1+\delta }|\alpha
\rangle _{2}+\sqrt{1-\delta }|-\alpha \rangle _{2}\right)   \notag \\
&&\left( \sqrt{1+\delta }|\alpha \rangle _{3}+\sqrt{1-\delta }|-\alpha
\rangle _{3}\right) \frac{1}{\sqrt{2}}\left( \sqrt{1+\delta }|g\rangle +%
\sqrt{1-\delta }|e\rangle \right) ,
\end{eqnarray}%
where $\widetilde{N}_{\alpha }=1/\sqrt{2+2\sqrt{1-\delta ^{2}}%
e^{-2\left\vert \alpha \right\vert ^{2}}}.$ For this case, we numerically
plot Fig. 6, which illustrates that the fidelity decreases with increasing
of $\delta $. Nevertheless, for $\delta \in \lbrack -0.1,0.1],$\ i.e., a $%
10\%$\ error in the weights of $|\alpha \rangle $\ and $|-\alpha \rangle $\
states as well as in the weights of $|g\rangle $\ and $|e\rangle $, a
fidelity greater than $91.56\%,92.38\%,$and $93.37\%$ can be achieved for $%
\kappa ^{-1}=45~\mu $s, $60~\mu $s, $100~\mu $s, respectively.

\begin{figure}[tbp]
\begin{center}
\includegraphics[width=10.0 cm, clip]{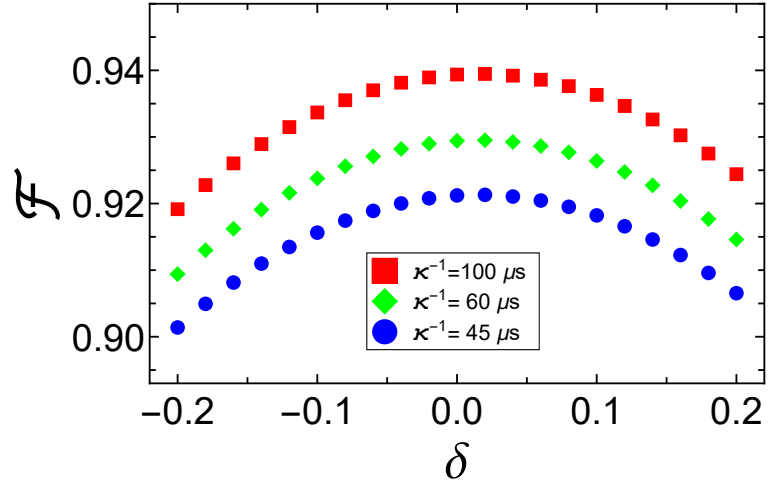} \vspace*{-0.08in}
\end{center}
\caption{(color online) Fidelity versus $\protect\delta $ for $\protect%
\kappa ^{-1}=45~\protect\mu $s, $60~\protect\mu $s, and $100~\protect\mu $s.}
\label{fig:6}
\end{figure}

\begin{figure}[tbp]
\begin{center}
\includegraphics[width=10.0 cm, clip]{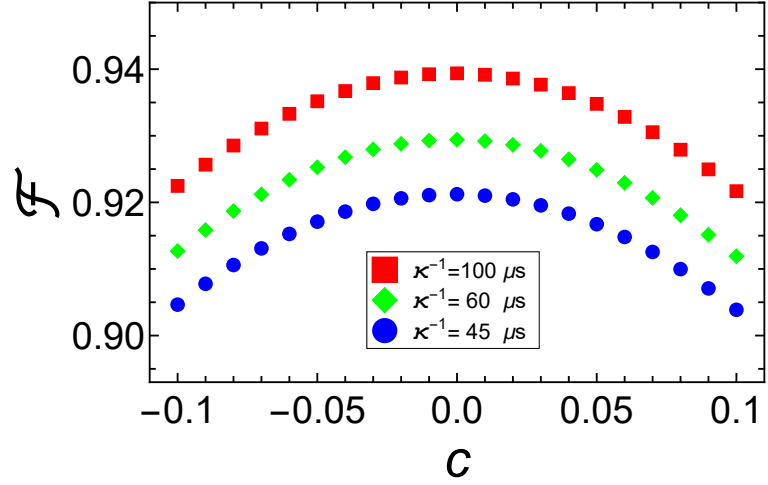} \vspace*{-0.08in}
\end{center}
\caption{(color online) Fidelity versus $c$ for $\protect\kappa ^{-1}=45~%
\protect\mu $s, $60~\protect\mu $s, and $100~\protect\mu $s.}
\label{fig:7}
\end{figure}

It may be an experimental challenge to have the condition (7) well
satisfied. Thus, to be more realistic, we modify the condition (7) (with $n=3
$\ for the three-cavity case)\ as follows:
\begin{equation}
\bigskip g_{1}^{2}/\Delta _{1}=\left( 1+c\right) g_{2}^{2}/\Delta
_{2}=\left( 1-c\right) g_{3}^{2}/\Delta _{3}.
\end{equation}%
We numerically plot Fig. 7, which illustrates the fidelity versus $c.$\ The
coupling constants $g_{2}$\ and $g_{3},$\ used in the numerical simulation
for plotting Fig. 7, are dependent of $c$\ and calculated according to
Eq.~(21), given the values of $g_{1},\Delta _{1},\Delta _{2},$and $\Delta
_{3}$\ listed in Table 1. Figure 7 shows that the fidelity decreases with
increasing of $c.$\ However, for $c\in \lbrack -0.05,0.05]$, i.e., a $5\%$\
deviation with respect to the ideal condition (7), the fidelity is greater
than $91.67\%,92.49\%,$and $93.48\%$\ for $\kappa ^{-1}=45~\mu $s, $60~\mu $%
s, $100~\mu $s, respectively. This result indicates that a high fidelity can
still be obtained when the condition (7) is not well satisfied.

The operational time for the GHZ state preparation is estimated to be $\sim
0.74$\ $\mu $s, which is much shorter than the decoherence time of the
transmon qutrit and the cavity decay time used in the numerical simulations.
For the cavity frequencies given in Table 1 and $\kappa ^{-1}=45~\mu $s, the
quality factors of the three cavities are $Q_{1}\sim 9.16\times 10^{5}$, $%
Q_{2}\sim 9.07\times 10^{5}$, and $Q_{3}\sim 8.99\times 10^{5}$, which are
available because a 1D microwave cavity or resonator with a high quality
factor $Q\gtrsim 10^{6}$\ was reported in experiments [37-42].

\begin{center}
\textbf{C. Discussion}
\end{center}

The analysis presented above shows that the operational fidelity is
sensitive to the error in the initial state preparation and the deviation
from the ideal condition (7). Our numerical simulations demonstrate that a
high fidelity can still be obtained when the error in the initial state
preparation and the deviation from the condition (7) are small. In order to
achieve a high fidelity, one would need to reduce the error in the initial
state prepare, reduce the deviation from the condition (7), select cavities
with a high quality factor, and choose the qutrit with a long coherence
time. One can also improve the fidelity by employing a coupler qutrit with a
larger level anharmonicity, so that the unwanted couplings between each
cavity and the irrelevant level transitions of the qutrit are negligible.
Finally, it should be remarked that further investigation is needed for each
particular experimental setup. However, this requires a rather lengthy and
complex analysis, which is beyond the scope of this theoretical work.

\begin{center}
\textbf{VI. CONCLUSIONS}
\end{center}

We have proposed a one-step method to realize a \textit{hybrid}
controlled-NOT gate with one SC qubit simultaneously controlling multiple
target cat-state qubits. This proposal operates essentially through the
dispersive coupling of each cavity with the coupler qutrit. To the best of
our knowledge, this work is the first to demonstrate the realization of the
proposed hybrid multi-target-qubit gate based on cavity or circuit QED. This
proposal is quite general and can be applied to implement a hybrid
controlled-NOT gate with one matter qubit (of different type) simultaneously
controlling multiple\ target cat-state qubits in a wide range of physical
systems.

As shown above, this proposal has the following features and advantages: (i)
The gate implementation is quite simple, because it requires only a
single-step operation; (ii) Neither classical pulse nor measurement is
needed; (iii) The hardware resources are significantly reduced because only
one coupler SC qutrit is needed to couple all of the cavities; (iv) The
intermediate higher energy level of the SC qutrit is not occupied during the
gate realization, thus decoherence from the coupler SC qutrit is
significantly reduced; and (v) The gate operation time is independent of the
number of target qubits, thus it does not increase with an increasing number
of target qubits.

As an application, we further discuss how to create a hybrid GHZ entangled
state of SC qubits and cat-state qubits, by applying the proposed hybrid
multi-qubit gate. To the best of our knowledge, our work is the first to
show the preparation of a \textit{hybrid} GHZ entangled state of SC qubits
and cat-state qubits. We have also numerically analyzed the experimental
feasibility of generating a hybrid GHZ state with one SC qubit and three
cat-state qubits within current circuit QED technology. We hope that this
work will stimulate experimental activities in the near future.

\begin{center}
\textbf{ACKNOWLEDGEMENTS}
\end{center}

This work was partly supported by the National Natural Science Foundation of
China (NSFC) (11374083, 11774076, U21A20436), the Jiangxi Natural Science
Foundation (20192ACBL20051), and the Key-Area Research and Development
Program of GuangDong province (2018B030326001).


\begin{thebibliography}{999}
\bibitem{s111} For a transmon qutrit, the $|g\rangle \leftrightarrow
|f\rangle $ transition is much weaker than those of the $|g\rangle
\leftrightarrow |e\rangle $ and $|e\rangle \leftrightarrow |f\rangle $
transitions. Thus, we have $\gamma _{fg}^{-1}\gg \gamma _{eg}^{-1},\gamma
_{fe}^{-1}.$
\end{thebibliography}
\end{document}